\documentclass[pra,twocolumn,amsmath]{revtex4}

\begin{document}

\title{Reply to ``Comment on ``Some implications of the quantum nature of laser fields for quantum computations'''' }

\author{Julio Gea-Banacloche}
\email[]{jgeabana@uark.edu}
\affiliation{Department of Physics, University of Arkansas, Fayetteville, AR 72701}

\date{\today}

\begin{abstract}
In this revised reply to quant-ph/0211165, I address the question of the validity of my results in greater detail, by comparing my predictions to those of the Silberfarb-Deutsch model, and I deal at greater length with the beam area paradox.   As before, I conclude that my previous results are an (order-of-magnitude) accurate estimate of the error probability introduced in quantum logical operations by the quantum nature of the laser field.  While this error will typically (for a paraxial beam) be smaller than the total error due to spontaneous emission, a unified treatment of both effects reveals that they lead to formally similar constraints on the minimum number of photons per pulse required to perform an operation with a given accuracy; these constraints agree with those I have derived elsewhere.
\end{abstract}
\maketitle

\section{Introduction}

In a recent comment \cite{comment}, Itano claims that my conclusions in a couple of recent papers \cite{pra,prl} are ``invalid'' because I use an ``inappropriate model'' (which he calls a reverse micromaser model) to describe the interaction between an atom in free space and a quantized laser pulse.  He also argues that for such a situation the decoherence rate is simply that given by spontaneous emission (presumably, as computed in the absence of the external field).  I address these two points in what follows.

\section{Errors arising from the quantum nature of the laser pulse}

Itano's first claim is, I think, highly exaggerated.  It is not like I am actually claiming to see micromaser dynamics, such as collapses or revivals, in free space; at most, I'm just using ``micromaser'' dynamics for the duration of one Rabi oscillation.  And I am not trying to get ``exact'' results to within precise numerical factors; the purpose of my calculation (in Section 2 of \cite{pra}), besides serving to motivate the rest of the paper, is just to derive {\it approximate\/} expressions for the order of magnitude, and the scaling, of the effects I am interested in.  

Moreover, these effects---specifically, the fact that there will, unavoidably, be entanglement in energy between the atom and the field in the case of a $\pi/2$ pulse, and the fact that the quantum field will exhibit fluctuations in both phase and amplitude---are so basic that they have to be present in any sensible treatment; they do not depend, in any essential way, on the most distinctive feature of micromaser, or Jaynes-Cummings, dynamics---namely, the fact that any emitted photon ``hangs around'' the atom and can be reabsorbed at any time. 

It seems to me that Itano needs to decide whether he is questioning the validity of the precise numerical factors I derive, or the whole scaling laws and order of magnitude estimates.  If the former, of course, I never claimed that kind of precision in the first place.  If the latter, he is simply wrong, as one can show from a very simple calculation based on the recent model developed by Silberfarb and Deutsch \cite{silberfarb} to analyze the interaction of a quantized laser pulse with an atom in free space. Their basic result, namely, their equation (19) (reproduced below)
\begin{equation}
  \frac{ d \hat{\rho}}{d t}
  =\frac{-i}{\hbar} \left[ H_{\text{coh}},\hat{\rho} \right]
    -\frac{\kappa}{2} \left\{ \hat \sigma _+\hat \sigma _-,\hat{\rho}  \right\}
    +\kappa \,\hat \sigma _- \hat{\rho} \hat \sigma _+ 
\end{equation}
is essentially equivalent to the Mollow model which Itano favors in his Comment: after a unitary transformation, the atom is driven by a classical field, and interacts with a set of quantized vacuum modes.  The only two things that Silberfarb and Deutsch have added to this picture are: first, they have already explicitly made the Markov approximation (equivalent to the Weisskopf-Wigner treatment of spontaneous emission), so that an emitted photon has no chance of coming back to the atom (which means that one could not expect to recover Jaynes-Cummings-type dynamics from this model, exactly, in any limit); and, second, they have only included those vacuum modes that can be considered to be traveling with the incident field in the first place (that is, the modes that are initially ``occupied,'' before the unitary transformation), which is why equation (1) has a modified ``atomic decay rate'' $\kappa$, which differs from the total vacuum spontaneous emission rate $\Gamma$ by what is, essentially, a geometrical factor (more on this below).

It is a trivial matter to integrate Eq.~(1), for instance, for a $\pi$ pulse, that is, for a time $T=\pi/2g\alpha$ (see \cite{silberfarb} for notation). Assuming an atom initially in the ground state $|b\rangle$, I get, for the probability to find it in the excited state $|a\rangle$ at the time $t$, 
\begin{equation}
\rho_{aa}(T) = 1-\frac{3\pi}{8}\,\frac{\kappa}{2 g\alpha}
\label{rho}
\end{equation}
that is to say, a ``failure probability,'' $p$, of $3\pi \kappa/16 g \alpha$ for this operation (a bit flip).  Substituting values from \cite{silberfarb}, and using $T=\pi/2g\alpha$, this becomes
\begin{equation}
p = \frac{3\pi^2}{32}\,\frac{\hbar\omega}{I A T} =  0.93 \frac{\hbar\omega}{P T} = \frac{0.93}{\bar n}
\label{p}
\end{equation}
where $I$ is the intensity, $A$ the mode cross-section, $P=I A$ the power, and $\bar n = P T/\hbar\omega$ the average number of photons in the pulse.  This is to be compared with the value $p=0.62/\bar n$ which I estimated in \cite{pra}, from an effective single mode approach, and which I attributed, in that picture, to (quantum) amplitude fluctuations in the laser field.

Of course, other ``gates'' yield different results, which also depend on the initial atomic state.  For instance, for a $\pi/2$ pulse (Hadamard transform), starting from the ground state, integration of (1) yields a (surprisingly low) failure probability of the order of $0.04/\bar n$, but starting from the excited state this becomes instead $0.43/\bar n$.  In \cite{pra}, I simply argued that $0.25/\bar n$ could be a reasonable estimate of the error probability due to the intrinsic phase fluctuations in a coherent state, and worked from there.  Clearly, if all of the above is any indication, I am unlikely to have missed the mark in any significant way.

Note, in particular, that all of the above results display, when expressed in terms of the mode area, the paradoxical behavior that Itano points out in his Comment, namely, ``the odd consequence that one can decrease the effective decoherence of the atomic system by increasing the cross-sectional area of the laser beam, keeping the intensity at the position of the atom fixed, no matter how distant the outer parts of the beam are from the atom, since doing so increases $\langle n \rangle$.'' In the previous version of this reply, I dismissed this effect a little too casually.  Now I think there is  a very simple explanation, in this particular picture. 

In the ``effective single mode'' picture, of course, what happens as one increases the area is just that the number of photons increases and the field accordingly becomes more classical, but there is a certain nonlocality to this interpretation which, as Itano points out, makes it somewhat unappealing.  In the framework of the Silberfarb-Deutsch model, on the other hand, the quantum nature of the field is contained only in the vacuum modes responsible for the decay rate $\kappa = \Gamma\sigma_{\text{eff}}/A$, and what happens as the beam becomes wider is simply that its direction of propagation becomes better and better defined, and the fraction of vacuum modes associated with it becomes smaller and smaller, relative to the total number of vacuum modes available to the atom.  In other words, the probability for the atom to emit spontaneously along the beam direction goes down, by a simple geometric factor, just because the ``acceptance angle'' goes down.  Hence, the decoherence {\it associated with the laser field\/}  really does go down as the beam becomes wider; equivalently, the field really {\it does\/} become more ``classical'' in this limit.

Of course, since the total number of vacuum modes available to the atom is always the same, one could also say that, as the cross-section of the beam increases, the decoherence associated with the laser field becomes a smaller and smaller fraction of the {\it total\/} decoherence, due to field quantization, experienced by the atom.  I explore this in the next Section. 

\section{Total error probability arising from field quantization (including empty modes)}

The observation at the end of the previous Section naturally leads into the second part of Itano's argument, namely, his claim that the laser field does not really introduce any decoherence beyond that already present from spontaneous emission.  This cannot be literally true, if by spontaneous emission is really meant the decay in the absence of any external field; as van Enk and Kimble \cite{vanenk2} have pointed out already, one cannot really ignore the interplay between the ``classical'' field and its quantized ``vacuum'' modes.

Nonetheless, it does seem natural to expect that the spontaneous emission rate in empty space will, in effect, set the order of magnitude of the ultimate decoherence rate due to the quantum nature of the electromagnetic field (see the Appendix for more thoughts along these lines).  In any case, in the framework of the present model---that is, under the Markov approximation and as long as we are only interested in the state of the atom---this total decoherence is easily calculated: one only has to replace $\kappa$ by $\Gamma$ in Eq.~(1), in order to include all the vacuum modes, and not only those associated with the laser field.

When this is done, one finds, for instance, for a $\pi$ pulse a total error probability of $3\pi \Gamma/16 g \alpha$.  Since the only difference between $\Gamma$ and $\kappa$ is the geometrical factor $A/\sigma_{\text{eff}}$, this means that Eq.~(\ref{p}) becomes
\begin{equation}
p = \frac{3\pi^2}{32}\,\frac{\hbar\omega}{I \sigma_{\text{eff}} T} =  = \frac{0.93}{\bar n'}
\label{pprime}
\end{equation} 
where $\sigma_{\text{eff}} =  3\pi/2 k^2$ is the cross section for scattering out of the paraxial modes, and $\bar n'$ is the number of photons in the volume $\sigma_{\text{eff}}\, cT$.  Note that, since the beam cannot really be focused to less than a wavelength, one will always have $A > \sigma_{\text{eff}}$, and hence the total number of photons in the pulse, $\bar n$, is always greater than $\bar n'$.  Thus, if one needs to keep the error probability per logical operation below a certain tolerance level $\epsilon$, the constraint $\bar n' > 1/\epsilon$ always implies $\bar n > 1/\epsilon$, which is my original result.  Thus, if anything, one could argue that my estimates of a minimum energy requirement are, for this kind of situation, overly conservative (more on this below, in the next Section \cite{vanenk}).

The advantage of the result (\ref{pprime}) is that it includes, for the case of an atom in free space manipulated by a laser pulse, all the decoherence effects arising from the quantum nature of the field (not just the laser field, but also all the unoccupied vacuum modes), and it shows explicitly that it is the number of photons within the volume $V = \sigma_{\text{eff}} \,c T$ that counts.  More precisely, one can show, also from the above equations, that what is needed is a minimum energy density, in the vicinity of the atom, of the order of $\hbar/\epsilon T$ per wavelength cubed. Obviously, while this does not contradict my earlier claims that the minimum number of photons in the pulse must be $\ge 1/\epsilon$, it does extend them and makes them more precise. 

\section{Discussion}

In conclusion, contrary to Itano's claims, it appears that my earlier estimates \cite{pra,prl} for the minimum energy requirements for quantum computation are not ``invalid,'' only too conservative; specifically, for the case of atoms being manipulated by paraxial laser pulses, the actual energy requirements will typically be larger (perhaps even much larger, if $A\gg \sigma_{\text{eff}}$) than what I have calculated.

I have no problem with that, nor can I really say that I am surprised.  My estimates are explicitly based on the assumption that the only quantum noise affecting the qubit is in the quantum fluctuations carried by what I call the ``control system'' itself: thus, in the case of a laser pulse, the quantum field fluctuations associated with the modes {\it that make up that pulse\/}.  In practice, an atom in free space will be coupled to many more quantized, empty, modes, and will become entangled with them, and hence experience decoherence, because of the possibility of spontaneous emission into those modes.  It goes without saying that such {\it additional\/} decoherence can only make matters worse.

To put it another way, it is implicitly assumed in my calculations that, in order to get close to the lower bound, one should have near-optimal coupling between the qubit and the control field, such as one might obtain in a microcavity.  Conversely, for any suboptimal coupling, such as that of an atom to a paraxial beam, I am only estimating a fraction of the total decoherence due to field quantization.  But that's all right.  I never claimed that I was trying to estimate the {\it whole\/} decoherence rate, only the decoherence that may be attributed to the interaction with the quantized control system.  This represents an absolute floor below which one cannot drop, no matter how hard one works to minimize other decoherence sources (for instance, in the case of an atom, by confining it to a microcavity).

In short, I would claim that all my results, as expressed in the form of inequalities, are (order-of-magnitude) correct, and that I have calculated, in my various publications \cite{pra,prl}, precisely what I said I was going to calculate.  

Having said this, I should add that, when one looks beyond order of magnitude estimates and starts paying attention to the actual numerical coefficients, one does find differences between the various models, and it would be interesting to understand better the reasons for these discrepancies; personally, one thing that all this has left me wondering is whether it is possible to ascertain really the roles played by amplitude and phase fluctuations, separately, within a proper, fully quantized, multimode treatment of the laser interaction with the atom.  These are interesting questions for which I would like to have an answer, even if they do not ultimately affect the main conclusions of my work.

\begin{acknowledgments}
I am actually grateful to Wayne Itano, for forcing me to think hard about all these things.  I have not communicated directly with A. Silberfarb and I. H. Deutsch, but I am grateful to them, too, for the very opportune posting of their paper (and for explicitly stating in it that my conclusions were correct!).
\end{acknowledgments}

\appendix*

\section{The minimum energy constraint from spontaneous emission}

In this appendix I provide what I think is the simplest (model-independent)  derivation of a minimum energy constraint based on spontaneous emission, and show that it holds even for far-detuned Raman transitions.

Let the spontaneous emission rate be $\Gamma$, and hence assume that atomic state purity (alternatively, information) is disappearing at this rate.  Let the time needed to perform a logical operation on the atomic qubit be $T$.  Then, if $\epsilon$ is the tolerable error rate (for instance, the fault-tolerant error-correction threshold), we clearly require
\begin{equation}
\Gamma  T < \epsilon 
\label{one}
\end{equation}
which means that one needs to apply a sufficiently strong field to the atom, since $T$ will be inversely proportional to the Rabi frequency $\Omega_R$.  Let's assume, for the sake of argument, that $T=\pi/\Omega_R$.  Then (\ref{one}) can be rewritten as
\begin{equation}
\frac{\pi^2\Gamma}{\Omega_R^2 T} < \epsilon 
\label{two}
\end{equation}
or, using standard expressions for $\Gamma$ and $\Omega_R$,
\begin{equation}
\frac{\pi^2}{4 \pi\epsilon_0}\,\frac{4\omega^3 d_{ab}^2}{3\hbar c^3}\,\left( \frac{d_{ab} E_0}{\hbar} \right)^{-2} < \epsilon T 
\label{three}
\end{equation}
(where $d_{ab}$ is the atomic dipole moment and $E_0$ the field amplitude).  Now use $\omega = 2 \pi c/\lambda$ and note that the e.m. field's energy density is $\frac{1}{2} \epsilon_0 E_0^2$.  One gets the constraint
\begin{equation}
\frac{1}{2} \epsilon_0 E_0^2 \, (\sigma_{\text{eff}}\, cT)  > \frac{\pi^2}{4} \, \frac{\hbar\omega}{\epsilon} 
\label{four}
\end{equation}
where $\sigma_{\text{eff}} = 3\pi/2 k^2$ is the effective cross-section for scattering out of the paraxial modes, as in the text.  In words, the number of photons within a volume $V$ of cross-sectional area $\sigma_{\text{eff}}$ and length $cT$ (the length of the pulse) has to be greater than $1/\epsilon$.  
This is (in order of magnitude) the same as the result (\ref{pprime}) derived in the text.

Equation (\ref{four}) also holds for the case, treated in \cite{pra}, of a transition driven by detuned Raman lasers.  In that case, it can be argued that the atom only spends a time of the order of $1/\Delta$ in the excited (intermediate) state, and hence the loss of purity due to spontaneous emission would be limited to $\Gamma /\Delta$, regardless of $T$; but, for large detuning, one has also an effective two-photon Rabi frequency of the order of $\Omega_R^2/\Delta$, so by combining $\Gamma/\Delta < \epsilon$ with $\Omega_R^2 T/\Delta = \pi$, and eliminating $\Delta$, one immediately obtains Eq.~(\ref{two}), and the rest proceeds as above.


\begin{thebibliography}
\bibitem{}
\bibitem{comment} W. Itano, quant-ph/0211165.
\bibitem{pra} J. Gea-Banacloche, Phys. Rev. A {\bf 65}, 022308 (2002).
\bibitem{prl} J. Gea-Banacloche, Phys. Rev. Lett. {\bf 89}, 217901 (2002).
\bibitem{silberfarb} A. Silberfarb and I. H. Deutsch, quant-ph/0210056.
\bibitem{vanenk} The observation that the decoherence due to entanglement with the laser field would typically be smaller than that due to spontaneous emission had already been made by S.J. van Enk and H. J. Kimble, J. Quantum Info. Comput. {\bf 2}, 1 (2002).
\bibitem{vanenk2} S. J. van Enk and H. J. Kimble, quant-ph/0212028.



\end{thebibliography}
\end{document}